\documentclass[a4 paper, 12 pt] {article}
\begin{document}

\title{\bf{On (2+1)Dimensional Topologically Massive Non-linear Electrodynamics}}
\author{M. \'{S}lusarczyk $^{a,b)}$ \thanks{mslus@phys.ualberta.ca} \, and A.
Wereszczy\'{n}ski $^{a)}$ \thanks{wereszcz@alphas.if.uj.edu.pl}
       \\
       \\ $^{a)}$ Institute of Physics,  Jagiellonian University,
       \\ Reymonta 4, Krakow, Poland
       \\
       \\  $^{b)}$ Department of Physics, University of Alberta,
       \\ Edmonton, AB T6G 2J1, Canada}
\maketitle

\begin{abstract}
The (2+1) dimensional non-linear
electrodynamics, the so called Pagels--Tomboulis electrodynamics,
with the Chern--Simons term is considered. We obtain "generalized
self--dual equation" and find the corresponding generalized
massive Chern--Simons Lagrangian. Similar results
 for (2+1) massive dilaton electrodynamics have been obtained.
\end{abstract}

\section{\bf{The Pagels--Tomboulis model}}
Among many gauge theories the Pagels--Tomboulis model \cite{Pagels}
\begin{equation}
L= -\frac{1}{4} \left( \frac{F_{\mu \nu }^a F^{a \mu \nu
}}{\Lambda^4} \right)^{\delta-1} F_{\mu \nu }^a F^{a \mu \nu }
\label{pagels}
\end{equation}
has a special place. Here $$F_{\mu \nu }^a=\partial_{\mu } A_{\nu
}^a -\partial_{\nu } A_{\mu }^a -\epsilon^{abc} A^b_{\mu }
A^c_{\nu }$$ is the standard field tensor, $\delta $ is a
dimensionless parameter and $\Lambda $ is a dimensional constant.
The gauge field is $SU(2)$ type i.e. $a=1,2,3$. One can check that
this Lagrangian has positively defined energy for $\delta \geq
\frac{1}{2}$.
\par
The Pagels--Tomboulis theory was originally proposed as an effective
model for the low energy (3+1) QCD \cite{Pagels}. In fact, it was shown
that in the frame of this model electrical sources are confined for
$\delta \geq \frac{3}{2}$.
Energy of the electric field generated by the external charge is
infinite due to the divergence at large distance. Moreover, the
dipole-like external source gives a finite energy field
configuration. The energy $\mathcal{E}$ behaves like
\begin{equation}
\mathcal{E} = c_0 |q|^{\frac{2 \delta }{2\delta -1 }}
\Lambda^{\frac{4\delta -4}{2\delta -1}} R^{\frac{2\delta
-3}{2\delta -1}}, \label{energypagels}
\end{equation}
where $c_0$ is a numerical constant, $q$ is an external charge and
$R$ is the distance between charges \cite{My1}. Interestingly enough
for $\delta = \frac{5}{2}$ we obtained the
$\sqrt{R}$ behaviour of the energy, which is in agreement with a
phenomenological potential found in fits to spectra of heavy
quarkonia \cite{kacper1}. The standard linear potential appears in
the limit $\delta \rightarrow \infty $. Highly non-linear gauge
models appear in studying of the (2+1) QCD as well. Corresponding
confining force has been recently obtained in \cite{gies}.
\par
The Pagels-Tomboulis model has been also considered
as an example of a field theory with the vanishing trace
of the energy-momentum tensor. In case of any $(n+1)$-dimensional
gauge theory defined by a Lagrangian $$L=L(F),$$ where $F=F^a_{\mu
\nu } F^{a \mu \nu }$, the trace $T = T_{\mu}^{\; \mu}$ has the following form
\begin{equation}
T= 4 \frac{ d L}{d F} F - (n+1) L.
\label{trace}
\end{equation}
One can easily find that for any $(n+1)$-dimensional space--time
there exists the unique Lagrangian (up to a
 multiplicative constant)
,
which gives the vanishing trace of the energy--momentum tensor
\begin{equation}
L= -\frac{1}{4}(F^a_{\mu \nu } F^{a \mu \nu })^{\frac{n+1}{4}}.
\label{tracelag}
\end{equation}
Only in $(3+1)$-dimensional space--time such a  Lagrangian is a linear
function of $F$. For instance, in $(2+1)$ dimension the pertinent
model takes the form
\begin{equation}
L_{2+1}=-\frac{1}{4} (F^a_{\mu \nu } F^{a \mu \nu
})^{\frac{3}{4}}. \label{lag2}
\end{equation}
Such particular Lagrangian, in its Abelian version, have been recently used
as a source in the Einstein equations. Many static spherically
symmetric solutions have been obtained \cite{Garcia}.
\par
This short and incomplete list of applications of the Pagels--Tomboulis model
in various areas of theoretical physics shows that the model is
very interesting and has rich mathematical structure.
Unfortunately, in contradistinction to other non-linear gauge theories
(for example the Born--Infeld theory \cite{born}) it has not been
considered in the systematic way.
\newline
In the present paper we focus on the (2+1) Abelian
Pagels--Tomboulis model with the additional topological term -
the Chern--Simons term
\begin{equation}
L= -\frac{1}{4} (F_{\mu \nu }F^{\mu \nu } )^{\delta} +\frac{m}{4}
\epsilon^{\mu \nu \rho } A_{\mu } F_{\nu \rho }.
\label{massive}
\end{equation}
Here, for simplicity, the dimensional constant $\Lambda $ has been
neglected. This model is the natural generalization of the
non--linear electrodynamics (\ref{lag2}) considered in
\cite{Garcia}. It is well known that the Chern--Simons part of the
Lagrangian (\ref{massive}) does not enter explicitly to the
expression for the energy. It is due to the fact that this term is
metric independent. Thus the energy--momentum tensor remains unchanged
in comparison with the pure non-linear electrodynamics case. It
has been shown using the field equations that in the Maxwell limit i.e.
for $\delta =1$ the gauge field from (\ref{massive}) is
proportional to the dual strength tensor \cite{Jackiw},
\cite{Deser}
\begin{equation}
A_{\mu }=\frac{1}{2m} \epsilon_{\mu \nu \rho} F^{\nu \rho}.
\label{maxdual}
\end{equation}
Of course, using the $U(1)$ gauge transformation $A_{\mu }
\rightarrow A_{\mu } +\partial_{\mu } \psi $ one can generate the
whole family gauge equivalent solutions. The solution
(\ref{maxdual}) corresponds to the Lorentz gauge. This self--dual
equation can be derived also from the massive Chern--Simons
Lagrangian \cite{Townsend}
\begin{equation}
L_{mass}=\frac{1}{m^2} A_{\mu }A^{\mu } -\frac{m}{4} \epsilon^{\mu \nu
\rho } A_{\mu } F_{\nu \rho}.
\label{maxproca}
\end{equation}
In fact, it was shown that these Lagrangians are equivalent.

Let us now generalize these results for all $\delta > \frac{1}{2}$.
The pertinent equations of motion read
\begin{equation}
\partial_{\nu } \left[ (F_{\sigma \rho }F^{\sigma \rho } )^{\delta -1} F^{\nu \mu
} \right] +\frac{m}{2 \delta } \epsilon^{\mu \nu \rho } F_{\nu \rho } =0.
\label{eqmot1}
\end{equation}
The solution of the second order equations (\ref{eqmot1}) has the
generalized form of the self-dual equation (\ref{maxdual})
\begin{equation}
A_{\mu } = \frac{\delta }{2m} (F_{\sigma \lambda }F^{\sigma \lambda }
)^{\delta -1} \epsilon^{\mu \nu
\rho } F_{\nu \rho }.
\label{selfdual}
\end{equation}
It is immediately seen that after differentiation of both side of
generalized self--dual equation and multiplication by $\epsilon^{\alpha
\beta \gamma}$ we obtain (\ref{eqmot1}). As in the Maxwell case, the
generalized self--dual equation emerges as a field equation from
generalized massive Chern--Simons Lagrangian
\begin{equation}
L_{mass}=\frac{1}{4} (f_{\mu} f^{\mu} )^{\frac{\delta }{2 \delta -1}} -
\frac{D \delta }{2\delta -1}
\epsilon^{\mu \nu \rho } f_{\mu } \partial_{\nu } f_{\rho },
\label{dual}
\end{equation}
where the $U(1)$ gauge field in the generalized
generalized massive Chern--Simons is denoted by $f_\mu$ to distinguish  it from
the corresponding field in the original Pagels--Tomboulis Lagrangian.
The field equations for (\ref{dual}) have the following form
\begin{equation}
\epsilon^{\mu \nu \rho } \partial_{\nu } f_{\rho } - \frac{1}{4D}
(f_{\nu} f^{\nu} )^{\frac{1-\delta }{2 \delta -1}} f^{\mu }=0.
\label{eqmot2}
\end{equation}
In order to establish the generalized self--dual equation for the new gauge
field $f_{\mu }$ one has to rewrite (\ref{eqmot2}) as
\begin{equation}
f_{\mu \nu } f^{\mu \nu } = \frac{1}{8D^2} (f_{\nu} f^{\nu} )^{\frac{1}{2
\delta -1}}.
\label{f}
\end{equation}
Here $f_{\mu \nu }= \partial_{\mu } f_{\nu } -\partial_{\nu } f_{\mu }$.
Then we express $f_{\mu } f^{\mu }$ in terms of the corresponding field
strength tensor and substitute this into the field equation
(\ref{eqmot2}). One eventually gets
\begin{equation}
f_{\mu } =2 \cdot 8^{\delta -1} D^{2
\delta -1}
\left(f_{ \nu  \rho} f^{\nu \rho } \right)^{\delta -1} \epsilon^{\mu
\lambda \sigma } f_{\lambda \sigma }.
\label{selfdual2}
\end{equation}
This generalized self-dual equation becomes identical to (\ref{selfdual})
if the constant $D$ reads
\begin{equation}
D=\left( \frac{2\delta }{m8^{\delta }} \right)^{\frac{1}{2\delta -1}}.
\label{const}
\end{equation}
One can see that equations (\ref{dual}) and (\ref{selfdual2}) are in
agreement with the results presented in \cite{Ilha}, where the case of
$\delta =\frac{p}{q}$, $p,q \in Z$ was considered. The relation between
the topological massive Pagels--Tomboulis model and the generalized
massive Chern--Simons model (\ref{dual}) becomes clearly visible when we
observe that they have the common origin, that is they follow from a
single Lagrangian
\begin{equation}
L_M=\frac{\alpha}{4} (f_{\mu} f^{\mu} )^{\frac{\delta }{2 \delta -1}}
-\beta
\epsilon^{\mu \nu \rho} f_{\mu } \partial_{\nu }A_{\rho } +\frac{m}{2}
\epsilon^{\mu \nu \rho} A_{\mu }\partial_{\nu } A_{\rho }
\label{master}
\end{equation}
where the constants are
\begin{equation}
\frac{\alpha }{\delta } \left( \frac{2 \beta^2}{\alpha^2}
\right)^{\delta }=1, \; \; \frac{\beta^2}{\alpha } = 2m \left(
\frac{2\delta }{m 8^{\delta }} \right)^{\frac{1}{2\delta -1}},
\label{masterconst}
\end{equation}
and the fields $A_{\mu }$ and $f_{\mu }$ are treated independently.
Indeed, after variation of (\ref{master}) with respect to $f_{\mu }$ one
can use the resulting equation to eliminate this field from the Lagrangian and
get (\ref{massive}). In the same way the gauge field $A_{\mu }$ can be
expressed in terms of $f_{\mu }$. As a result we get the generalized
massive Chern--Simons model. The Lagrangian (\ref{master}) gives in the
limit $\delta =1$ the so called master Lagrangian \cite{Jackiw}.
\section{\bf{The dilaton model}}
Let us now find the analogous dual structure for the dilaton-like
Lagrangian (it is possible to add a potential term for the scalar field
but it does not change the result obtained below)
\begin{equation}
L=-\frac{\sigma (\phi )}{4} F_{\mu \nu } F^{\mu \nu } +\frac{m}{4}
\epsilon^{\mu \nu \rho } A_{\mu } F_{\nu \rho} +\frac{1}{2}
(\partial_{\mu } \phi )^2
\label{dylaton}
\end{equation}
In fact, as it was shown in \cite{Pagels} the models
(\ref{massive}) and (\ref{dylaton}) share many features
(especially in the context of the low energy QCD where the
topological term is omitted). It emerges from the fact that they
can be understood as the usual electrodynamics in rather an
unusual medium. In the other words both models have the form $L=
\epsilon F_{\mu \nu}F^{\mu \nu}+...$ where the dielectric function
$\epsilon $ is a function of $ F_{\mu \nu}F^{\mu \nu}$ in the
Pagels--Tomboulis model or $\phi $ in the dilaton model. In
particular, in (\ref{dylaton}) $\sigma( \phi)=\phi^{\delta -1}$
plays the same role as $\epsilon (F_{\mu \nu }F^{\mu \nu })=
(F_{\mu \nu }F^{\mu \nu })^{\delta -1} $ in (\ref{massive}) (see
e.g. \cite{My1}, \cite{My2}).
\par
On the other hand, the Lagrangian (\ref{dylaton}) appears in the
natural way as a part of the topological generalization of the
$(2+1)$ dilaton--Maxwell--Einstein theory \cite{dilaton}. This
theory has been treated as the toy model of the quantum
gravitation. There have been found exact solutions describing the
formation of a black hole by collapsing matter. The Hawking
radiation can be also described in the frame of this model. The
particular form of $\sigma $ function is
motivated by the string theory and usually reads
$$\sigma = e^{a \phi }, $$ where $a$ is a dimensionless constant.
However, some other forms of $\sigma $ have been also under
consideration \cite{dilatoninne}.
\newline
The equations of motion are as follows
\begin{equation}
\partial_{\mu } (\sigma  F^{\mu \nu }) +\frac{m}{2} \epsilon^{\nu \mu \rho}
F_{\mu \rho}=0
\label{eqmot3}
\end{equation}
and
\begin{equation}
\partial_{\mu } \partial^{\mu } \phi + \frac{1}{4} \sigma' F_{\mu \nu } F^{\mu \nu
}=0,
\label{eqmot4}
\end{equation}
where prime denotes the differentiation with respect to the scalar field.
It is easy to notice that the solution of the equation (\ref{eqmot3}) has
self--dual-like form
\begin{equation}
A_{\mu }= \frac{\sigma (\phi ) }{2m} \epsilon^{\mu \nu \rho} F_{\nu \rho
}.
\label{selfdualdyl}
\end{equation}
The corresponding massive Chern--Simons-like Lagrangian is found to be
\begin{equation}
L_{mass} =\frac{m^2}{2} \cdot \frac{1}{\sigma (\phi )} f_{\mu } f^{\mu }
-\frac{m}{2} \epsilon^{\mu \nu \rho } f_{\mu } \partial_{\nu } f_{\rho}
+\frac{1}{2} (\partial_{\mu } \phi )^2.
\label{dylatondual}
\end{equation}
The pertinent field equations read
\begin{equation}
f_{\mu } = \frac{1}{2 m } \sigma (\phi ) \epsilon^{\mu \nu \rho } f_{\nu
\rho }
\label{selfdualdyl2}
\end{equation}
and
\begin{equation}
\partial_{\mu } \partial^{\mu } \phi + \frac{m^2}{2} \cdot
\frac{\sigma'}{\sigma^2} f_{\mu } f^{\mu }=0.
\label{eqmot5}
\end{equation}
It is immediately seen that the self--dual equation (\ref{selfdualdyl})
for the massive Chern--Simons--dilaton model (\ref{dylatondual}) is just the
equation of motion. Moreover, using (\ref{selfdualdyl2}) one eliminates the
field $f_{\mu }$ from the second field equation. After that equation
(\ref{eqmot5}) takes the from
\begin{equation}
\partial_{\mu } \partial^{\mu } \phi +\frac{1}{4} \sigma' f_{\mu \nu } f^{\mu \nu
}=0.
\label{eqmot6}
\end{equation}
As we have expected both Lagrangian (\ref{dylaton}) and
(\ref{dylatondual}) give the same equation of motion. Additionally
we see mutual duality of these models. The strong coupling sector
of the one theory is interchanged with the weak coupling sector in
the other one.
\newline
At least at the theoretical level one can consider a model where the
dielectric function depends on $U(1)$ gauge invariant $F_{\mu \nu }F^{\mu
\nu }$ as well as on the scalar function $\phi $:
\begin{equation}
L= -\frac{\sigma}{4} (F_{\mu \nu }F^{\mu \nu } )^{\delta} +\frac{m}{4}
\epsilon^{\mu \nu \rho } A_{\mu } F_{\nu \rho } +\frac{1}{2} (\partial_{\mu } \phi
)^2.
\label{massivedyl}
\end{equation}
We see that the Pagels--Tomboulis and the dilaton model are included in
this Lagrangian and can be derived in the particular limits. It is easy
to check that the corresponding generalized self-dual equation has the
form
\begin{equation}
A_{\mu } = \frac{\delta \sigma (\phi ) }{2m} \epsilon^{\mu \nu \rho }
F_{\nu
\rho } (F_{\sigma \lambda }F^{\sigma \lambda } )^{\delta -1}
,
\label{dualmassivedyl}
\end{equation}
whereas the massive Chern--Simons-like Lagrangian
\begin{equation}
L_{mass}= A \sigma^{\frac{1}{1-2\delta }} (f_{\mu} f^{\mu}
)^{\frac{\delta }{2
\delta -1}} -
\frac{D \delta }{2\delta -1}
\epsilon^{\mu \nu \rho } f_{\mu } \partial_{\nu } f_{\rho }
+\frac{1}{2} (\partial_{\mu } \phi )^2.
\label{procamassivedyl}
\end{equation}
Here the constants read
\begin{equation}
\frac{D}{A}=\left(\frac{\delta 2^{\delta }}{2m}
\right)^{\frac{1}{1+2\delta }}, \; \; \;
A \left(\frac{D}{A} \right)^{2 \delta }=2^{\delta -2} (2\delta -1)
\label{const2}
\end{equation}
\section{\bf{Conclusions}}
In the present paper we have considered the (2+1) Pagels--Tomboulis
electrodynamics with topological term. The generalized version
of the self-dual equations and the corresponding massive
Chern--Simons-like Lagrangian have been found. Moreover, we have
proved that both models can be derived from the generalized master
equation (\ref{master}).
\par
The dual structure has been also obtained in case of the (2+1)
topological dilaton--Maxwell model. There are two equivalent
Lagrangians (\ref{dylaton}) and (\ref{dylatondual}) consisting of
scalar field and $U(1)$ field. It seems to be interesting that the
strong coupling regime in the first theory is related to the weak
coupling sector in the second. The non--perturbative effects in
one model can be reformulated as the perturbative effects in the
other one and solved applying standard methods. Knowing that
$(2+1)$ topological dilaton--Maxwell model plays an important role
in studying $(2+1)$ gravity we believe that this feature can give us
possibility to find some new gravitational solutions for the modified model.
It is quite remarkable that the dual structure can be found not only in $U(1)$
gauge models. Field theories containing additional degrees of
freedom (here the scalar field) possess the dual formulation as
well. The problem whether such a dual structure is observed in
case of more complicated additional field is still unsolved and
requires separate studies.
\par
Similar duality has been observed in the combined
Pagels--Tomboulis--dilaton model.
\par
There are two obvious directions in which the present work can be
continued. First of all, as it was mentioned before, the full
$(2+1)$ topological dilaton--Maxwell--Einstein theory should be
considered. Secondly, because of the fact that the
Pagels--Tomboulis Lagrangian is mostly considered in its
non--Abelian version it seems to be important to analyze the
non--Abelian generalization of the results obtained here. Then the
topological term takes the form of the well-known $SU(2)$ Chern--Simons
invariant. Very interesting, new results, concerning the standard
$\delta =1$ case, have been recently obtained \cite{wotzasek}.
\newline
\par
We would like to thank Professor H. Arod\'{z} for many helpful
comments and suggestions. This work was supported in part by the
ESF programme COSLAB.

\end{document}